\documentclass[twocolumn,aps,showpacs,floatfix,prc]{revtex4}
\usepackage[dvips]{epsfig}
\begin{document}

\title{ Relativistic kinematics for reactions involving massless particles }

\author{ D.N. Basu$^1$ and  Tapan Mukhopadhyay$^2$ }
\affiliation{Variable  Energy  Cyclotron  Centre, 1/AF Bidhan Nagar, Kolkata 700 064, India }

\email[E-mail 1: ]{dnb@veccal.ernet.in}
\email[E-mail 2: ]{tkm@veccal.ernet.in}

\date{\today }

\begin{abstract}

    Some useful kinematical relations for the absorption of a photon by a nucleus and its recoil are derived for the relativistic incident energies. These expressions provided for the relativistic kinematics of photoabsorption reactions, though simple, will be immensely useful for experimentalists as well as theoreticians.


\end{abstract}

\maketitle

\section{Introduction}
\label{section1}

    Experiments in which a photon ($\gamma$) with relativistic energy incident on a nucleus is absorbed by it, and the compound nucleus (the $\gamma$-nucleus system C*) subsequently undergoes nuclear reaction in one of the many available channels, have recently become the subject of a good deal of experimental interest, largely because of the availability of the highly energetic photons at the Saskatchewan Accelerator Laboratory (SAL) \cite{Sa99,Sa00}, and Jefferson Laboratory (JLab) \cite{Ce00,Ce01}. The excitation energy and relativistic kinematical relations in the reactions of the form $\gamma$ + C $\rightarrow$ C* are derived. Explicit expressions are provided for several quantities of interest and also conveniently expressed in the centre of mass system.

\section{Relativistic kinematics for the reactions of the form $\gamma$ + C $\rightarrow$ C* }
\label{section2}

    We consider a photon with kinetic energy $E_\gamma$ (as measured in the laboratory) in the relativistic domain is incident on a nucleus at rest in the laboratory with rest mass $m_0$ and derive expressions for several quantities of interest. 

\subsection{The laboratory system}

    The laboratory system is the frame of reference imagined to be fixed in the laboratory in which the experiment is being perfomed. In the reaction stated above, the nucleus is assumed to be the target which is at rest in this system and the $\gamma$ is incident on it with kinetic energy $E_\gamma$ as measured in this system. The energy conservation in the laboratory system is, therefore, given by

\begin{equation}
 [E_{\gamma} + m_0 c^2]^2 = m_0^{\prime 2} c^4 + p^2 c^2
\label{seqn1}
\end{equation}   
\noindent
where $m_0$ and $m_0^{\prime}$ are the rest masses of the nucleus before and after the photon absorption respectively, $p$ is the momentum of the recoiling nucleus in the laboratory frame and $c$ is the velocity of light in vacuum. The momentum conservation in the laboratory system can be easily written as 

\begin{equation}
 \frac{h \nu_0}{c} = \frac{E_{\gamma}}{c} = p
\label{seqn2}
\end{equation}   
\noindent
where $h$ is the Planck's constant and $\nu_0$ is the frequency of the incident photon in laboratory frame.

\subsubsection{Excitation energy of the nucleus}

    The excitation energy $E^*$ of the nucleus is the difference of rest mass energies of the nucleus before and after the photon absorption. Solving Eq.(1) for $m_0^{\prime}$ along with Eq.(2) one obtains 

\begin{equation}
 m_0^{\prime} = m_0 (1 + 2E_{\gamma}/m_0 c^2 )^{1/2} 
\label{seqn3}
\end{equation}   
\noindent
so that the excitation energy $E^*$ of the nucleus is given by 

\begin{eqnarray}
 E^*=&&m_0^{\prime}c^2-m_0 c^2 = m_0 c^2 [(1 + 2E_{\gamma}/m_0 c^2 )^{1/2} - 1] \nonumber \\
      =&&[m_0c^2(2E_{\gamma}+m_0c^2)]^{1/2}-m_0c^2
\label{seqn4}
\end{eqnarray}   

\subsubsection{Recoil velocity of the nucleus}

    After absorption of $\gamma$ with incident energy $E_\gamma$ (as measured in the laboratory frame), the nucleus initially with rest mass $m_0$ and now with rest mass $m_0^{\prime}$ recoils with a velocity $v_r$ as measured in the laboratory frame. Then the momentum $p$ of the recoiling nucleus in the laboratory frame is just

\begin{equation}
 p = m^{\prime} v_r = \frac{m_0^{\prime} v_r}{\sqrt{1- v_r^2/c^2}}
\label{seqn5}
\end{equation}   
\noindent
where $m^{\prime}$ is the moving mass of the recoiling nucleus in the laboratory frame. Substituting the value of $p$ from Eq.(2) and solving for the recoil velocity $v_r$ one obtains 

\begin{equation}
 v_r= \frac{E_{\gamma}c}{[E_{\gamma}+m_0 c^2]}
\label{seqn6}
\end{equation}   
\noindent
and the kinetic energy $E_r$ of the recoiling nucleus in the laboratory frame is

\begin{eqnarray}
 E_r=&&m^{\prime}c^2-m_0^{\prime}c^2 =\frac{m_0^{\prime}c^2}{\sqrt{1-v_r^2/c^2}}-m_0^{\prime}c^2 \nonumber \\
      =&&E_{\gamma}+m_0c^2-[m_0c^2(2E_{\gamma}+m_0c^2)]^{1/2}
\label{seqn7}
\end{eqnarray}   
\noindent
The kinetic energy $E_r$ of the recoiling nucleus in the laboratory frame can now be rewritten as the obvious result

\begin{equation}
\vspace{-0.16cm}
 E_r= E_{\gamma} - E^*.
\label{seqn8}
\end{equation}   

\subsection{The centre of mass system}

    The centre of mass system of interacting particles is defined as a frame of reference where the sum of the momenta of all interacting particles is zero. Therefore, in the centre of mass system incident photon and the target nucleus moves with momenta which are equal in magnitude $p_{cm}$ and opposite in direction. The photon momentum in the centre of mass frame can be obtained using relativistic Doppler effect whereas the momentum of the target nucleus in the centre of mass system is just $m v_{cm}$ where $m$ is the moving mass of the target nucleus in the centre of mass frame before collision and $v_{cm}$ is the velocity of the centre of mass of the $\gamma$-nucleus system measured in the laboratory system. Thus equating the momenta of the photon and the target nucleus before collision in the centre of mass frame 

\begin{equation}
 p_{cm} = \frac{\frac{h\nu_0}{c} [1 - v_{cm}/c]}{\sqrt{1- v_{cm}^2/c^2}} = m v_{cm} = \frac{m_0 v_{cm}}{\sqrt{1- v_{cm}^2/c^2}}
\label{seqn9}
\end{equation}   
\noindent
one obtains expression for the velocity $v_{cm}$ of the centre of mass of the $\gamma$-nucleus system measured in the laboratory system.

\begin{equation}
 v_{cm}= \frac{E_{\gamma}c}{[E_{\gamma}+m_0 c^2]}
\label{seqn10}
\end{equation}   
\noindent
which is also equal to the recoil velocity $v_r$ measured in the laboratory system. Thus 

\begin{equation}
 v_{cm}= v_r
\label{seqn11}
\end{equation}   
\noindent
The photon momentum and the momentum of the target nucleus in the centre of mass frame are equal in magnitude $p_{cm}$ and opposite in direction and is given by 

\begin{equation}
 p_{cm}= \frac{E_{\gamma}m_0c}{\sqrt{[m_0c^2(2E_{\gamma}+m_0 c^2)]}}
\label{seqn12}
\end{equation}   

    Similarly, the kinetic energy in the centre of mass frame $E_{cm}$ which is sum of the kinetic energies of the $\gamma$ and the nucleus moving in the centre of mass frame, is given by
 
\begin{eqnarray}
 E_{cm} =&& \frac{h\nu_0 [1 - v_{cm}/c]}{\sqrt{1- v_{cm}^2/c^2}}+[\frac{m_0 c^2}{\sqrt{1- v_{cm}^2/c^2}} - m_0c^2] \nonumber \\
 =&& [m_0c^2(2E_{\gamma}+m_0c^2)]^{1/2} - m_0c^2
\label{seqn13}
\end{eqnarray}   
\noindent
and in this case the kinetic energy $E_{cm}$ in the centre of mass frame is also equal to the excitation energy $E^*$. Thus 

\begin{equation}
 E_{cm}= E^*
\label{seqn14}
\end{equation}   

\section{Relativistic kinematics for $\gamma$ + $\gamma$ collisions }
\label{section3}

The case of two photons with energies $E_1$ and $E_2$ in laboratory frame of reference and moving in same direction is uninteresting as they will both move with the same velocity $c$ and will never collide and the centre of mass remains undefined. When they move in opposite directions, the velocity $v_{cm}$ of the centre of mass of the $\gamma$-$\gamma$ system measured in the laboratory system can be obtained by equating the momenta $p_{cm}$ of the two photons moving in opposite direction in the centre of mass system. Thus

\begin{equation}
 p_{cm} = \frac{\frac{h\nu_1}{c} [1 - v_{cm}/c]}{\sqrt{1- v_{cm}^2/c^2}} = \frac{\frac{h\nu_2}{c} [1 + v_{cm}/c]}{\sqrt{1- v_{cm}^2/c^2}}
\label{seqn15}
\end{equation}   
\noindent
where and $\nu_1 = E_1/h$ and $\nu_2 = E_2/h$ are the frequencies of photons in the laboratory frame. Therefore, solving Eq.(15) for $v_{cm}$ one obtains,

\begin{equation}
 v_{cm} = \frac{E_1 - E_2}{E_1 + E_2} c
\label{seqn16}
\end{equation}   
\noindent
and

\begin{equation}
 p_{cm} = \frac{\sqrt{E_1E_2}}{c}
\label{seqn17}
\end{equation}   
\noindent
where we have assumed $E_1 \ge E_2$. The kinetic energy in the centre of mass frame $E_{cm}$ which is sum of the kinetic energies of the two $\gamma$'s moving in the opposite direction in the centre of mass frame, is given by

\begin{equation}
 E_{cm} = \frac{h\nu_1 [1 - v_{cm}/c]}{\sqrt{1- v_{cm}^2/c^2}} + \frac{h\nu_2 [1 + v_{cm}/c]}{\sqrt{1- v_{cm}^2/c^2}} = 2 \sqrt{E_1E_2}.
\label{seqn18}
\end{equation}   
\noindent

\section{Summary and conclusion}
\label{section4}

    In summary, the excitation energy and kinematical relations in the reactions of the form  C +  $\gamma \rightarrow$ C* are derived relativistically. Expressions have been provided for several quantities of interest in the laboratory and the centre of mass frames of references including $\gamma$ + $\gamma$ collisions.

\end{document}